\documentclass[10pt,conference]{IEEEtran}

\usepackage{xcolor}

\usepackage{multirow} 

  \usepackage[pdftex]{graphicx}
\usepackage{algorithm}
\usepackage{algpseudocode}
\usepackage{amsmath}
\IEEEoverridecommandlockouts
\makeatletter
\def\ps@IEEEtitlepagestyle{
  	\def\@oddfoot{\Footer} 
	}
\let\old@ps@headings\ps@headings
\let\old@ps@IEEEtitlepagestyle\ps@IEEEtitlepagestyle
\def\confheader#1{%
  \def\ps@headings{%
    \old@ps@headings%
    \def\@oddhead{\strut\hfill#1\hfill\strut}%
    \def\@evenhead{\strut\hfill#1\hfill\strut}%
    \def\@oddfoot{\Footer} 
  }%
  \def\ps@IEEEtitlepagestyle{%
    \old@ps@IEEEtitlepagestyle%
    \def\@oddhead{\strut\hfill#1\hfill\strut}%
    \def\@evenhead{\strut\hfill#1\hfill\strut}%
  }%
  \ps@headings%
}
\makeatother
\confheader{} 
\def\Footer{
   {\footnotesize
  \begin{minipage}{\textwidth}
  \centering \scriptsize{DISTRIBUTION STATEMENT A. Approved for public release: distribution is unlimited. Approval ID: $<$XXXX-2025-XXXX$>$. (MARKED FOR DEMO PURPOSES ONLY)\\
  }  \end{minipage}
  }
}
\begin{document}
%
%
\title{Leveraging Photonic Interconnects for Scalable and Efficient Fully Homomorphic Encryption\\
\vspace{-0.2in}}
\author{\IEEEauthorblockN{Dewan Saiham, Di Wu, and Sazadur Rahman}
\IEEEauthorblockA{
Department of Electrical and Computer Engineering, University of Central Florida\\
\{dewan.saiham, di.wu, mohammad.rahman\}@ucf.edu
}
\vspace{-0.3in}
}

\maketitle
\begingroup
\renewcommand\thefootnote{}
\footnotetext{\textsuperscript{*}This is the author's version of the paper presented at GOMACTech 2025.}
\endgroup


\begin{abstract}
Fully Homomorphic Encryption (FHE) facilitates secure computations on encrypted data but imposes significant demands on memory bandwidth and computational power. While current FHE accelerators focus on optimizing computation, they often face bandwidth limitations that result in performance bottlenecks, particularly in memory-intensive operations. This paper presents \emph{OptoLink}, a scalable photonic interconnect architecture designed to address these bandwidth and latency challenges in FHE systems. \emph{OptoLink} achieves a throughput of 1.6 TB/s with 128 channels, providing 300 times the bandwidth of conventional electrical interconnects. The proposed architecture improves data throughput, scalability, and reduces latency, making it an effective solution for meeting the high memory and data transfer requirements of modern FHE accelerators.
\\
\end{abstract}
\renewcommand\IEEEkeywordsname{Keywords}

%

\section{Introduction}

Fully Homomorphic Encryption (FHE) represents a breakthrough in privacy-preserving computation, allowing encrypted data to be processed without revealing the plaintext. This capability is critical for safeguarding sensitive data in untrusted environments such as cloud computing, healthcare, and financial systems~\cite{marcolla2022survey}. By performing operations on encrypted inputs and returning encrypted outputs, FHE ensures robust security even in the event of server breaches, as the decryption key remains confidential. Key computational tasks in FHE schemes, including integer-based and ring learning with errors (R-LWE) methods, involve resource-intensive operations like large integer and polynomial multiplications~\cite{gentry2011implementing}. The Number Theoretic Transform (NTT), essential for modular polynomial multiplication, reduces complexity from $O(n^2)$ to $O(nlogn)$~\cite{fan2012somewhat}. However, implementing NTT is challenging due to high memory bandwidth requirements and complex data access patterns in hardware~\cite{cheon2017homomorphic}. Hardware acceleration using FPGA, ASIC, and Compute-in-Memory (CiM) platforms has improved efficiency but faces scalability limitations~\cite{roy2019fpga, reagen2021cheetah}. 

The high bandwidth demands and intricate memory access patterns of NTT often lead to read-after-write conflicts~\cite{riazi2020heax}, further exacerbated by the large parameters required for secure FHE~\cite{zhou2023fhemem}. NTT architectures that are pipelined and parallel have been proposed to increase efficiency~\cite{Duong-Ngo}, but they frequently lack programmability and adaptability across a range of security requirements. Effectively managing dataflow while resolving major memory bandwidth and access conflicts is a major challenge~\cite{zhang2022sok}. Although pipeline stalls have been employed to mitigate memory conflicts, the bandwidth requirements of large FHE parameters surpass the capabilities of traditional electrical interconnects. This necessitates novel interconnect solutions for enabling scalable and high-performance FHE systems.

\begin{figure}[!t]
\centering
\includegraphics[width=0.45\textwidth]{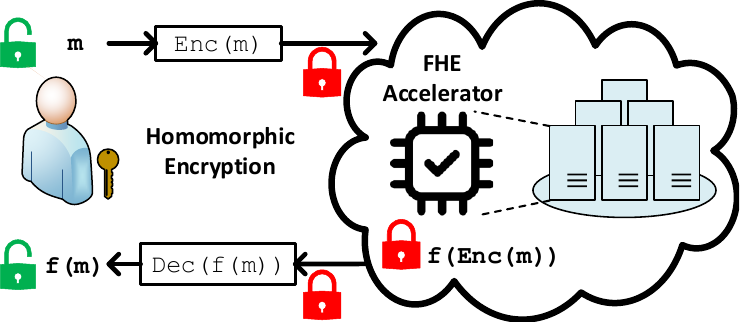}
\caption{Computational flow in fully holomorphic encryption (FHE).}
\label{fig:FHE}
\vspace{-0.25in}
\end{figure}

To address these challenges, we propose \emph{OptoLink}, a photonic interconnect architecture tailored for FHE applications. By replacing conventional electrical interconnects with optical alternatives, \emph{OptoLink} alleviates bandwidth bottlenecks and simplifies data paths in NTT computations. Utilizing technologies such as wavelength-division multiplexing (WDM) and space-division multiplexing (SDM), photonic interconnects support high-throughput, low-latency communication and scalable multi-chiplet designs. While effective in domains like deep neural networks~\cite{li2022spacx}, their potential in FHE architectures remains largely unexplored.

The primary contributions of this work include:
\begin{enumerate}
\item Identify memory bandwidth limitations as the key bottleneck in existing FHE accelerators, demonstrating that compute acceleration alone cannot ensure scalability.
\item We propose \emph{OptoLink}, a photonic interconnect architecture that reduces memory access conflicts and achieves high bandwidth for NTT operations. \emph{OptoLink} supports scalable deployments with data rates tailored to FHE requirements (Sec.\ref{sec:methodology}).
\item Utilizing photonics process design kits (PDKs) and electronic photonic design automation (EPDA) tools such as \texttt{Synopsys OptSim} and \texttt{OptoCompiler}, we design a scalable \emph{OptoLink} architecture achieving $1.6$TB/s bandwidth with $128$ optical channels, with potential for even higher throughput (Sec.\ref{subsec:simulation} and~\ref{sec:result}).
\end{enumerate}

The remainder of the paper is structured as follows: Sec.\ref{sec:motivation} outlines the background and motivation for \emph{OptoLink}. Sec.\ref{sec:methodology} details the proposed architecture and its implementation. Sec.\ref{sec:result} presents results and analysis, followed by conclusions in Sec.\ref{sec:conclusion}.

\section{Background and Motivation} \label{sec:motivation} 
\subsection{Number Theoretic Transform (NTT)}\label{subsec:NTT}
The NTT, a finite-field adaptation of the Fast Fourier Transform (FFT), enables efficient polynomial multiplication without roundoff errors involved with complex numbers. For a polynomial $a(x)= \sum^{n-1}_{i=0} a_ix^i$, the NTT is defined as:
\vspace{-0.1in}
\begin{equation}
\tilde{a_i}= \sum^{n-1}_{j=0} a_j \omega^{i.j} \; \mathrm{mod} \; q,
\label{e1}
\end{equation}
where $\omega$ is a primitive $n$-th root of unity in the ring $Z_q$, satisfying $\omega^n\equiv 1 \,mod\, q$, and $q$ is a prime number such that $q \equiv 1 \,mod\, n$. Polynomial multiplication using NTT involves transforming two polynomials $a$ and $b$ into the NTT domain, performing point-wise multiplication, and then applying the inverse NTT (INTT) to obtain the final result.

\begin{equation}
c= INTT(NTT(a) \circ NTT(b)),
\label{e2}
\end{equation}

Here $\circ$ denotes the element-wise multiplication. The INTT, which transforms data back from the NTT domain to its original form, is expressed as follows,
\vspace{-0.1in}
\begin{equation}
\vspace{-0.1in}
a_j = \frac{1}{n} \sum_{i=0}^{n-1} \tilde{a}_i \cdot \omega^{-i \cdot j} \,mod\ q
\label{e3}
\end{equation}
where, $\omega^{-i \cdot j}$ represents the inverse twiddle factor, and the scaling factor $n^{-1} \,mod\ q$ completes the transformation. This approach reduces the time complexity from $O(n^2)$ in naive polynomial multiplication to $O(n\,log\,n)$, similar to the FFT, but without floating-point precision errors. The Cooley-Tukey~\cite{cooley1965algorithm} and Gentleman-Sande~\cite{gentleman1966fast} algorithms are widely used for computing NTT and INTT efficiently, further optimizing the polynomial multiplication process by breaking it down to smaller subproblems through butterfly operations.

\subsection{Overview of current FHE Accelerators}\label{subsec:limitation}
FHE has made significant progress in lowering its original computational overhead, which was $10^9$ times slower than conventional unencrypted processing, since its inception in $2009$~\cite{gentry2009fully}. However, FHE operations are still $10,000 \times$ to $100,000 \times$ slower than traditional techniques, which makes them difficult to use in practice and emphasizes the need for specialized hardware accelerators. Due to their parallel processing capabilities, which may provide speedups of up to $257 \times$ when compared to CPUs~\cite{jung2021over}, GPUs have become a feasible solution. 
TensorFHE~\cite{fan2023tensorfhe} achieved efficiency levels comparable to ASIC-based systems, with a $1625.6 \times$ performance boost over CPUs and a $2.9 \times$ improvement over F1+. However, because GPUs are not made especially for FHE workloads, memory-intensive operations result in inefficiencies and significant power consumption. By enabling custom implementations of FHE tasks, such as the NTT, FPGAs offer increased flexibility. Significant performance improvements have been shown by accelerators like HEAX~\cite{riazi2020heax} and Poseidons~\cite{yang2023poseidon}, with Poseidon offering an improvement of more than $1000 \times$ over GPU solution. Additionally, designs like FAB optimize resource utilization to handle FHE operations effectively, showcasing the potential of FPGA-based acceleration~\cite{agrawal2023fab}. ASIC accelerators that are specifically made for FHE schemes, such as BFV and CKKS, perform much better. For example, ARK~\cite{kim2022ark} and CraterLake~\cite{samardzic2022craterlake} use innovations like hardware-accelerated bootstrapping and optimized data handling to solve performance bottlenecks and enable deeper computation depths, leading to significant speedups over GPU-based methods. Nevertheless, obstacles including their enormous device sizes, high power consumption, and substantial memory needs make it difficult for ASICs to be used practically and provide real-world adoption issues.

\begin{table}[htbp]
\caption{Memory Requirements of current FHE accelerators}
\begin{center}
\setlength\tabcolsep{10pt}
\begin{tabular}{|c|c|c|c|}
\hline
\textbf{Name} & \textbf{\textit{Hardware}} & \textbf{\textit{Supported}} & \textbf{\textit{Bandwidth}} \\
 & \textbf{\textit{Target}} & \textbf{\textit{Schemes}} & \\
\hline
100x~\cite{jung2021over} & GPU & BFV, & $900 \text{ GB/s}$ \\
 &  & CKKS & \\
\hline
cryptGPU~\cite{tan2021cryptgpu} & GPU & MPC & $1.25 \text{ GB/s}$ \\
\hline
TensorFHE~\cite{fan2023tensorfhe} & GPU & BFV, & $2.4 \text{ TB/s}$ \\
 &  & CKKS & \\
\hline
HEAX~\cite{riazi2020heax} & FPGA & CKKS & $34 \text{ GB/s}$ \\
 &  &  & $64 \text{ GB/s}$ \\
\hline
Poseidon~\cite{yang2023poseidon} & FPGA & BFV, & $460 \text{ GB/s}$ \\
 &  & CKKS & \\
\hline
FAB~\cite{agrawal2023fab} & FPGA & BFV, & $460 \text{ GB/s}$ \\
 &  & CKKS & \\
\hline
F1~\cite{samardzic2021f1} & ASIC & BFV, & $1 \text{ TB/s}$ \\
 &  & CKKS & \\
\hline
CraterLake~\cite{samardzic2022craterlake} & ASIC & BFV, & $2.4 \text{ TB/s}$ \\
 &  & CKKS & \\
\hline
BTS~\cite{kim2022bts} & ASIC & BFV, & $1 \text{ TB/s}$ \\
 &  & CKKS & \\
\hline
ARK~\cite{kim2022ark} & ASIC & BFV, & $1 \text{ TB/s}$ \\
 &  & CKKS & \\
\hline
\end{tabular}
\label{tab:fhe_survey}
\end{center}
\vspace{-0.25in}
\end{table}

\subsection{Memory Bottlenecks in FHE Acceleration}

Although there has been progress in computation acceleration, memory bandwidth is still a significant constraint in FHE applications~\cite{de2021does}. When compared to plaintexts, ciphertexts greatly increase the quantity of data, particularly in schemes like CKKS. This results in frequent memory accesses and severe bandwidth limitations. According to~\cite{kim2022ark}, a chip that has $40,960$ modular multipliers at $2~\text{GHz}$ and $3~\text{TB/s}$ HBM3 can do multiplications in $0.18~\text{ms}$, however it needs $2.1~\text{ms}$ for data transfer. This shows that in large-scale FHE jobs, the main bottleneck is not computing but data transportation. The intricate memory access patterns of FHE make bandwidth issues worse. Large memory allocations are needed for twiddle factors and intermediate results in operations like NTT, which frequently exceed on-chip cache capacity and call for frequent off-chip accesses. 
Static architectures find it difficult to handle the dynamic data dependencies and hardware resource strain caused by the $(n \log n)/2$ butterfly operations in FFT/NTT pipelines~\cite{zhang2022sok}.
Key-switching's resource-intensive actions increase memory needs even further. The decomposition parameter ($dnum$) impacts both memory and computation, requiring trade-offs between parallelization techniques such as residue-polynomial-level parallelism (rPLP) and coefficient-level parallelism (CLP). While global NTT communication increases latency in CLP, rPLP introduces additional data exchanges during basis conversion. Achieving hardware that dynamically balances these approaches remains a challenge. Furthermore, off-chip transfers of twiddle factors and intermediate results are frequently required due to limited on-chip memory, which exacerbates latency and power consumption~\cite{zhou2023fhemem}.

FHE accelerators need high-bandwidth capability to transport data efficiently because of these demands. State-of-the-art FHE accelerators have bandwidth requirements, which are highlighted in Table~\ref{tab:fhe_survey}. Meeting these demands presents considerable hurdles. \textbf{No electronic interconnects currently in use can achieve such high data transfer rates~\cite{de2021does}.}

\subsection{Overcoming Bandwidth Limits in FHE}\label{subsec:motivation}

To address bandwidth constraints, chiplet-based FHE accelerators utilize high-bandwidth memory (HBM) technologies like HBM3, offering up to $0.819~\text{TB/s}$ per stack with a 1024-bit data width~\cite{jedec_hbm3}. Advanced FHE accelerators (Table~\ref{tab:fhe_survey}) often require multiple HBM3 stacks to meet multi-terabyte bandwidth demands. Photonic interconnects, such as the $OptoLink$ architecture, provide superior bandwidth, achieving $0.8~\text{TB/s}$ with only $64$ channels, reducing bitwidth requirements by $16\times$ compared to HBM3, and scaling effectively for advanced workloads. $OptoLink$ dynamically multiplexes data, enabling flexible routing, task parallelism, and improved resource utilization with low latency. It supports diverse FHE tasks, including key-switching and NTT operations, without significant architectural changes. The design and experimental validation of $OptoLink$ are discussed in Secs.~\ref{subsec:optolink_arch} and \ref{sec:result}, highlighting its ability to meet bandwidth needs and improve scalability for privacy-preserving applications.

\section{Methodology} \label{sec:methodology}

\subsection{Photonic Interconnects}\label{subsec:photonic_interconnect}

\begin{figure}[!t]
\centering
\includegraphics[width=0.45\textwidth]{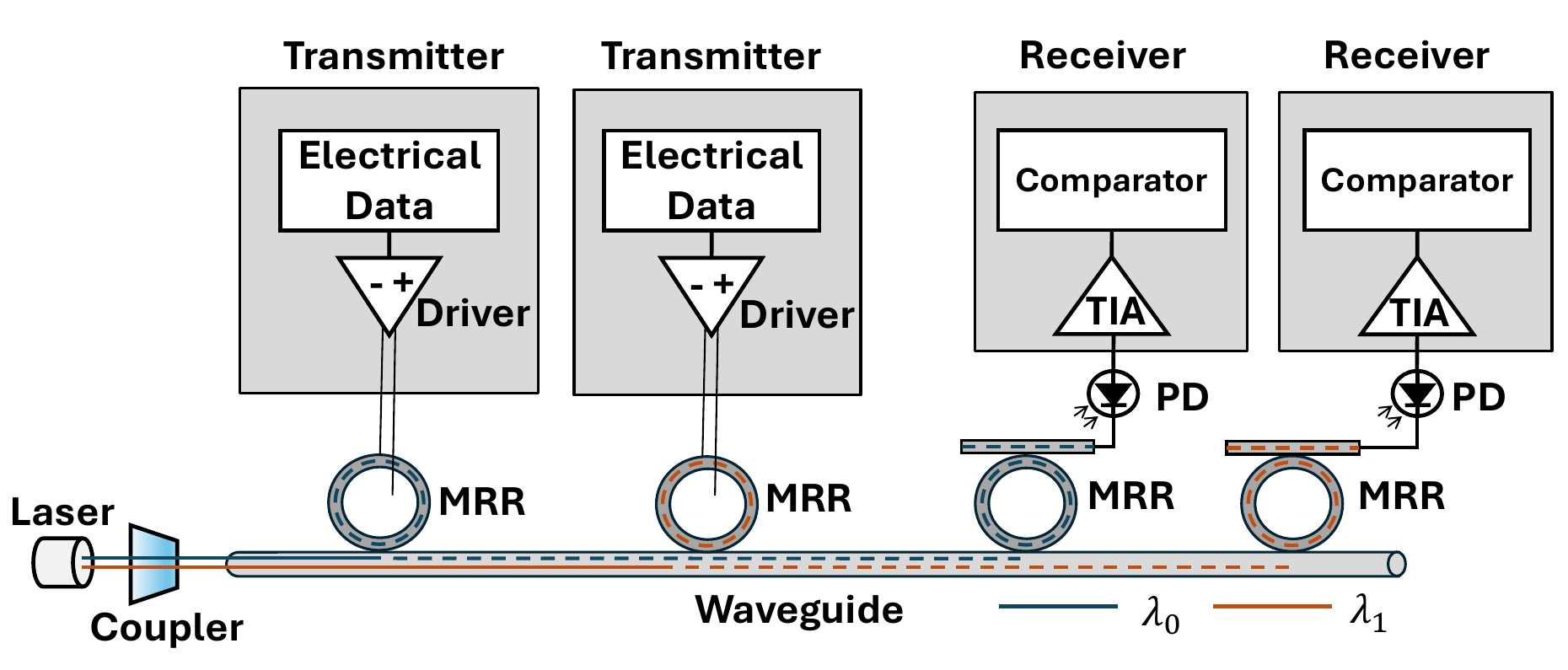}
\caption{Two transmitters and receivers are connected by a WDM photonic interconnect that operates on two distinct wavelengths, $\lambda_1$ and $\lambda_2$.}
\label{fig:interconnect}
\vspace{-0.2in}
\end{figure}

Photonic interconnects, which use light rather than conventional electrical signals, are a state-of-the-art method for high-speed data transfer in chip layouts. As depicted in Fig.~\ref{fig:interconnect}, light generated by an external laser is directed into on-chip waveguides through optical couplers. Micro-ring resonators (MRRs), serving as modulators and filters, are precisely tuned to specific wavelengths. These MRRs, equipped with resistive heaters and thermal tuning systems, maintain stability by compensating for process and thermal variations~\cite{Miller2009}. Electrical signals are modulated onto light by MRRs, with each signal assigned a unique wavelength. The modulated light propagates through waveguides to the receiver, where specific MRRs filter the signals. Photodetectors (PDs) then convert the optical signals back to electrical form, and transimpedance amplifiers (TIAs) amplify them for reliable data recovery.

To boost data throughput, WDM enables multiple data streams to transmit simultaneously on different wavelengths within a single waveguide. Advanced WDM systems can handle up to 64 wavelengths, each at $10$Gb/s, achieving aggregate bandwidths exceeding $100$Gb/s~\cite{werner2017designing}. Additionally, SDM further increases capacity by employing multiple parallel waveguides. By integrating WDM and SDM, photonic interconnects deliver exceptional bandwidth and energy efficiency. These attributes make them ideal for data-intensive applications like FHE accelerators, where efficient communication between cores and memory is critical.

\subsection{Single \emph{OptoLink} Channel} \label{subsec:architecture}

\begin{figure}[!t]
\centering
\includegraphics[width=0.47\textwidth]{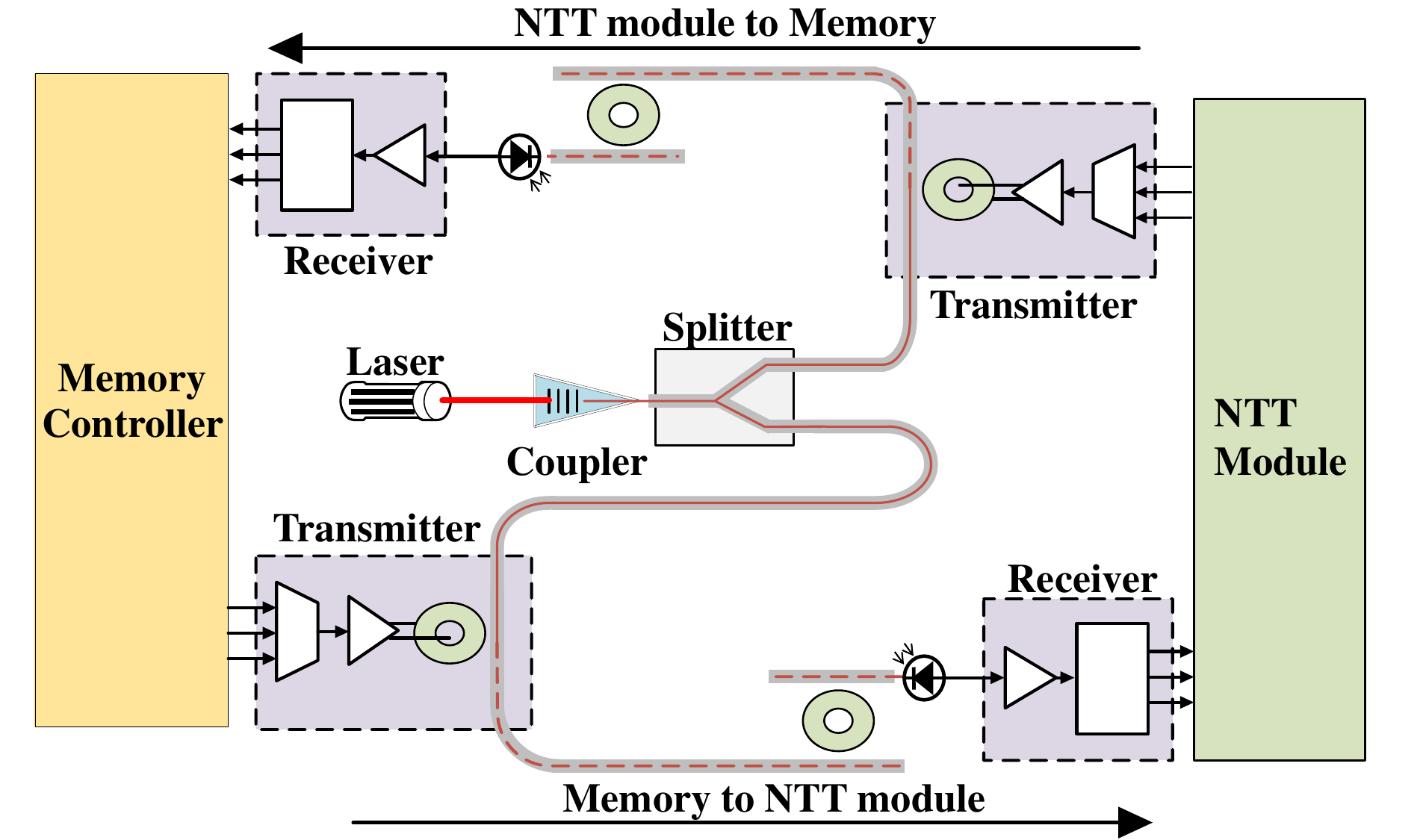}
\caption{Schematic of an WDM-enabled $OptoLink$ network communicating between the NTT module and memory. The NTT module retrieves input data and twiddle factors from memory, and sends the computed outputs back.}
\label{fig:interconnect_to_memory}
\vspace{-0.1in}
\end{figure}

Photonic interconnects are used to facilitate fast data transfer in Fig.~\ref{fig:interconnect_to_memory}, which illustrates the integration of memory and NTT modules within a single \emph{OptoLink} channel. Several signals are sent simultaneously via a single waveguide by the system using WDM, and each signal is given a distinct wavelength. This design substantially boosts bandwidth and ensures seamless communication between the memory controller and the NTT module. Memory is used to store input data close to the transmitters, including twiddle factors and coefficients. Using analog electrical signals derived from digital inputs, MRRs modulate light at certain wavelengths. The signals are isolated on the receiving end by wavelength-specific MRRs and sent to PDs for optical-to-electrical conversion. TIAs amplify these signals, which are then processed by comparators to recreate the original data. The output data undergoes similar modulation, transmission, and demodulation after processing by the NTT module. It is then sent back to the memory controller for subsequent operations. The same wavelengths are utilized for input and output data via different waveguides in order to maximize system efficiency. This reduces the overall number of wavelengths needed.

\subsection{Scalable \emph{OptoLink} Network Architecture}\label{subsec:optolink_arch}

\begin{figure}[!t]
\centering
\includegraphics[width=0.48\textwidth]{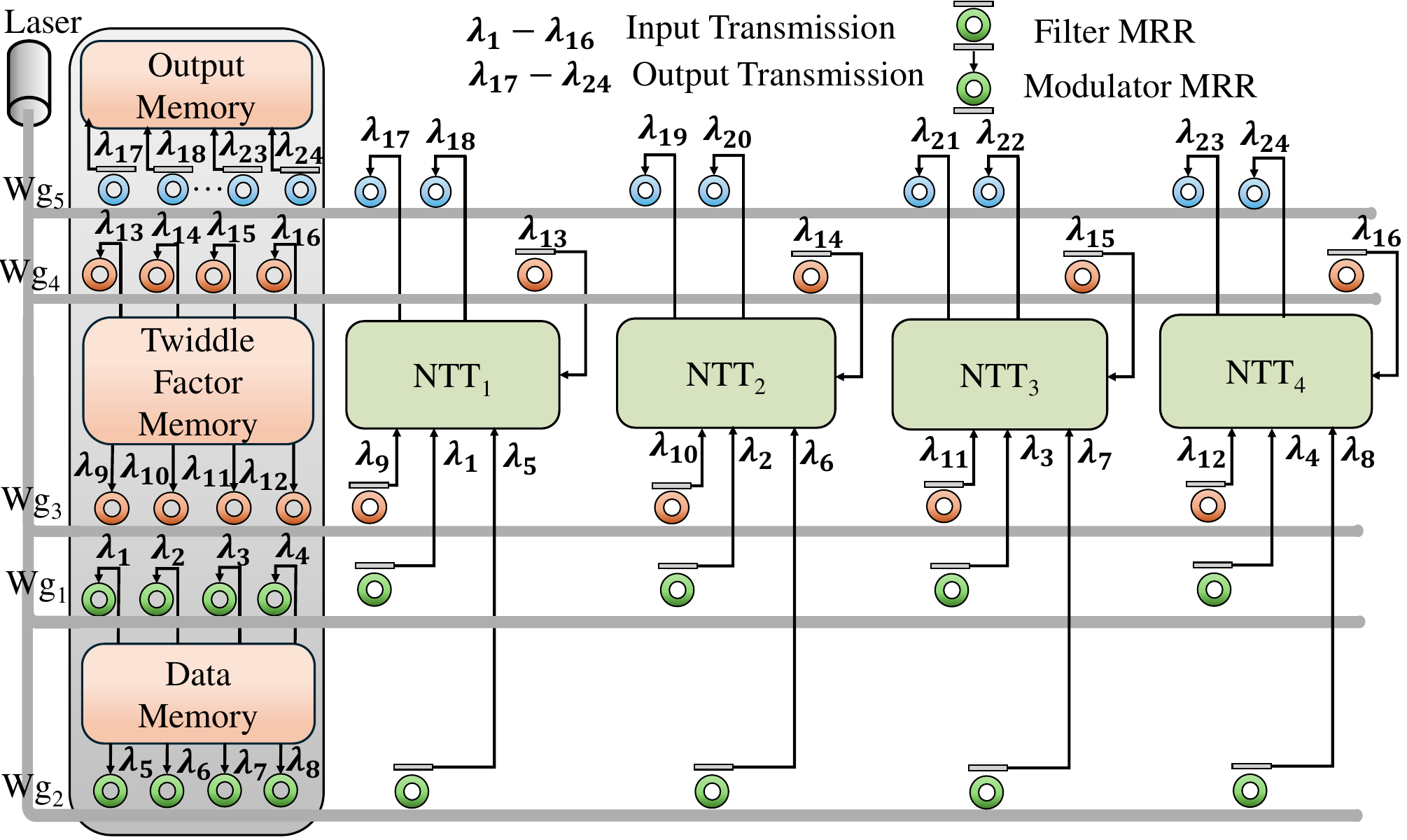}
\caption{Schematic representation of the $OptoLink$ architecture connecting four NTT modules via five waveguides. Wavelengths $\lambda_1 - \lambda_{16}$ are allocated for input data transmission, while $\lambda_{17} - \lambda_{24}$ handle output data transmission.}
\label{fig:optolink_to_NTT}
\vspace{-0.2in}
\end{figure}

The \emph{OptoLink} architecture, depicted in Fig.~\ref{fig:optolink_to_NTT}, integrates with four NTT modules and uses five waveguides for efficient data and twiddle factor transmission. \emph{Waveguides 1} and \emph{2} carry input data, to the NTT modules, while \emph{Waveguides 3} and \emph{4} deliver twiddle factors. Processed outputs are sent back to the memory controller via \emph{Waveguide 5}. Two wavelength groups facilitate communication: $\lambda_1 - \lambda_{16}$ transmit input data and twiddle factors, while $\lambda_{17} - \lambda_{24}$ handle results. Each wavelength corresponds to one bit per channel, with parallel optical channels enabling simultaneous data transmission. Wavelength reuse further enhances the system's scalability and efficiency.

Scalability, a critical requirement for FHE accelerators, is a key strength of \emph{OptoLink}. Increasing the number of optical channels and employing WDM significantly boost throughput while maintaining a compact physical footprint. Research shows that up to 64 wavelengths can be multiplexed within a single waveguide~\cite{werner2017designing}, providing substantial bandwidth expansion. However, this scalability introduces challenges in power consumption. Adding more NTT cores and optical channels increases laser power requirements to counteract insertion losses. While MRR tunability enables each transmitter to support multiple receivers, reducing the number of modulators, the cumulative power demand rises as more off-chip lasers, MRRs, and photodetectors are required. Balancing scalability with power and thermal efficiency is essential, particularly for high-throughput FHE workloads. The \emph{OptoLink} design effectively addresses bandwidth and latency constraints while offering flexibility for diverse applications. 

\subsection{Simulation Platform and Parameter}\label{subsec:simulation}

\begin{table}[t]
\caption{Photonic Parameters Utilized for Evaluation in \emph{OptoLink}}
\begin{center}
\setlength\tabcolsep{20pt}
\begin{tabular}{|c|c|}
\hline
\textbf{Component} & \textbf{\textit{Value}} \\
\hline
Laser Source & $5 \ dB$ \\
\hline
Coupler & $1 \ dB$ \\
\hline
Splitter & $0.2 \ dB$ \\
\hline
Waveguide & $1 \ dB/cm$ \\
\hline
Ring Drop & $0.7 \ dB$ \\
\hline
Ring Through & $0.01 \ dB$ \\
\hline
Photodetector & $0.5 \ dB$ \\
\hline
Ring Heating & $0.32 \ mW$ \\
\hline
\end{tabular}
\label{tab:params}
\end{center}
\vspace{-0.2in}
\end{table}

To assess the effectiveness of the proposed \emph{OptoLink} architecture, optical interconnects were implemented between NTT cores and off-chip memory to address the challenges posed by HEAX~\cite{riazi2020heax}, an FPGA-based FHE accelerator. HEAX’s complex memory-to-NTT module connections highlight the limitations of conventional electronic interconnects, motivating our adoption of photonic solutions for improved efficiency and reduced latency. Photonic parameters such as detector responsivity, modulator insertion loss, and coupling efficiency were included in the \texttt{Synopsys OptoCompiler} simulation of the \emph{OptoLink} design (Table~\ref{tab:params}). These factors helped determine the laser power needs, guaranteeing dependable signal delivery even in the face of optical imperfections. Furthermore, we used \texttt{Synopsys Design Compiler} to do time, power, and area analysis for the electrical network. To evaluate our architecture's scalability under various computing demands, the analysis took into account different numbers of NTT modules, notably configurations of 4, 8, and 16. 

\section{Results and Analysis} \label{sec:result}

\subsection{Timing Analysis}\label{subsec:timing}

\begin{figure}[!t]
\centering
\includegraphics[width=0.48\textwidth]{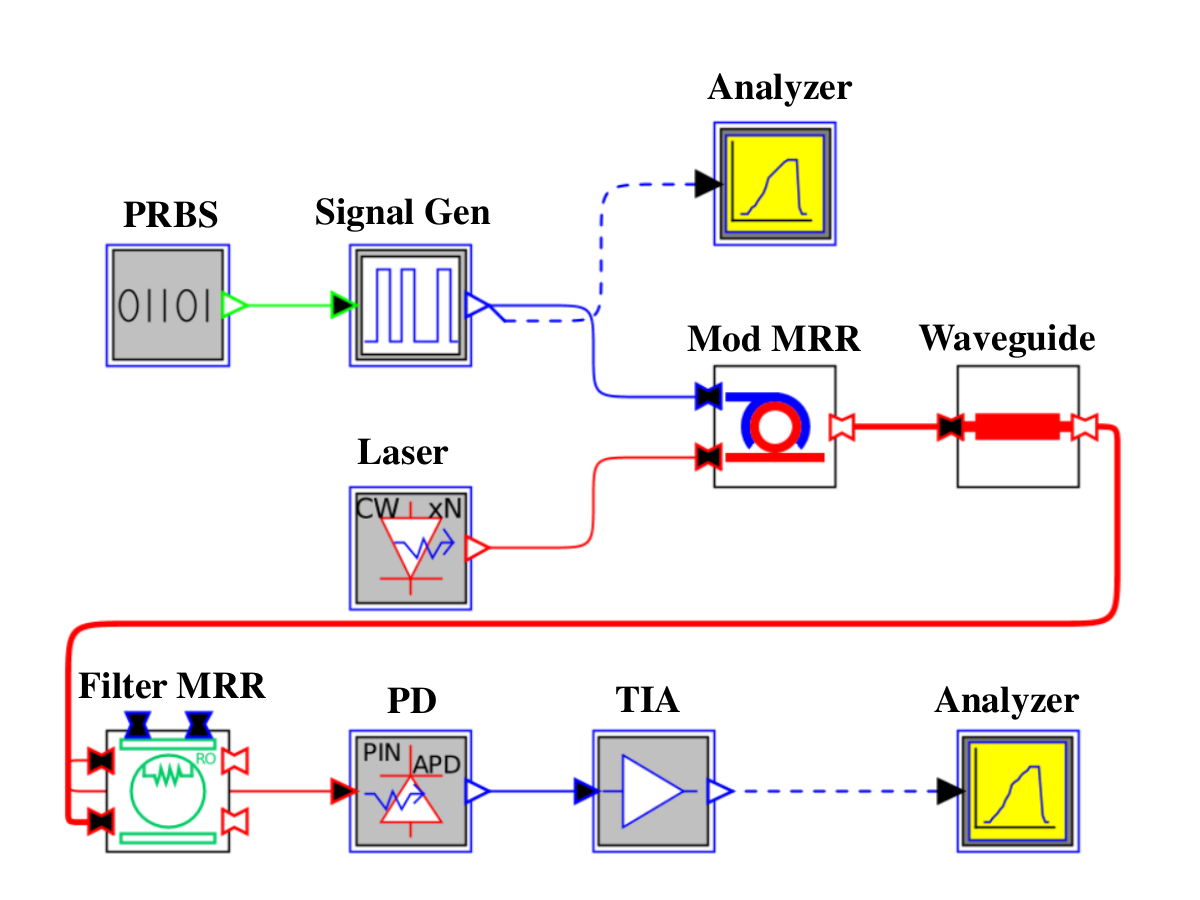}
\caption{Simulation configuration for a single channel in the \emph{OptoLink} system}
\label{fig:sim_setup}
\vspace{-0.2in}
\end{figure}

\begin{figure}[t]
\centering
\includegraphics[width=0.49\textwidth]{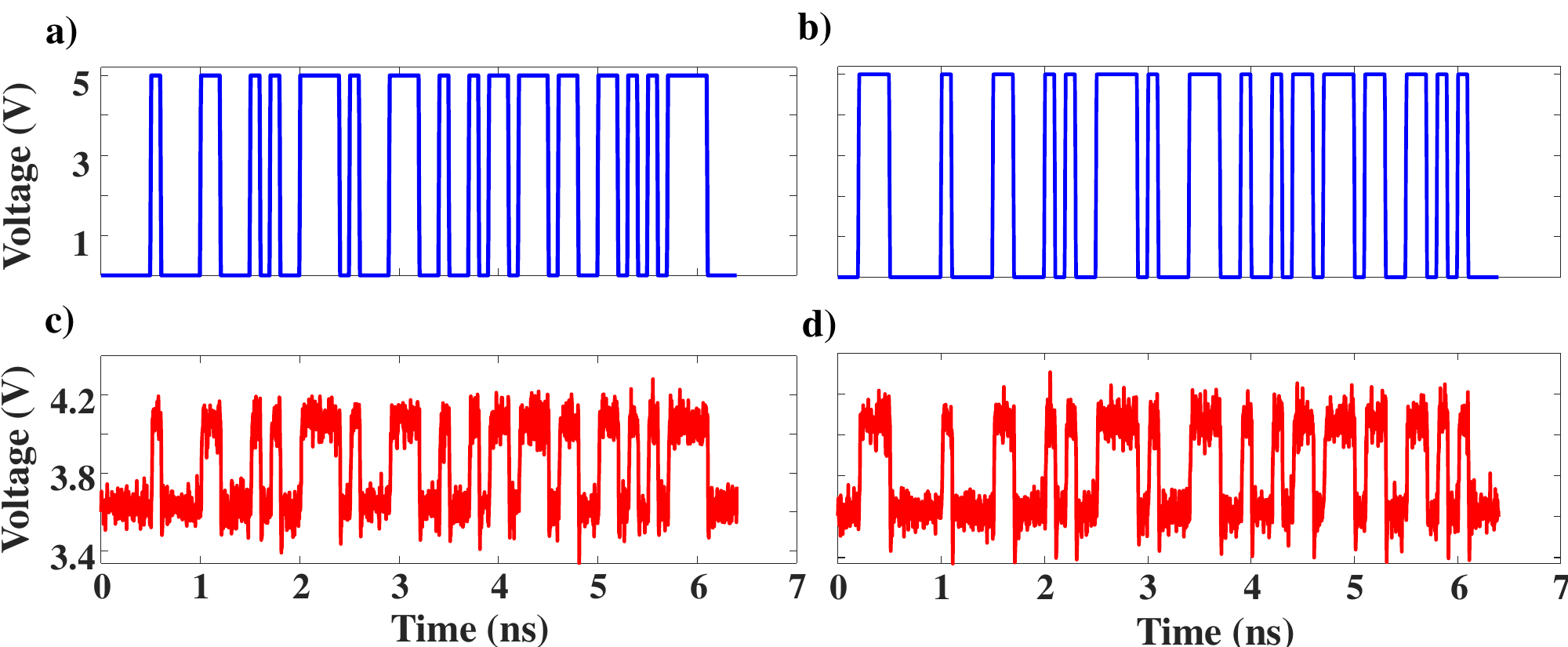}
\caption{{(a-b) Electrical input signals supplied to modulator MRRs across two separate $OptoLink$ channels for designated wavelengths. (c-d) Corresponding output signals, after optical-to-electrical conversion by the PD and amplification via the TIA in each channel, observed for the same wavelengths.}}
\label{fig:wdm_result}
\vspace{-0.2in}
\end{figure}

\texttt{Synopsys OptoCompiler} was used to simulate two optical channels in data transmission studies to assess the performance of the \emph{OptoLink} network. Data was sent at $10$Gb/s via a pseudo-random bit sequence (PRBS) generator, which needed $6.4$ns to produce a complete sequence. Fig.~\ref{fig:wdm_result} shows $64$-bit data sequences modulated by MRRs onto $1550$nm (channel 1) and $1551$nm (channel 2) wavelengths, with reliable signal recovery after transmission. Using a $1000\mu$m waveguide, \emph{OptoLink} achieved a transmission latency of $10$ps, significantly lower than the $3.04$ns required in an electrical network. Each \emph{OptoLink} channel achieved a data rate of $100$Gb/s or $12.5$GB/s, with a total bandwidth of $1.6$TB/s for 128 channels, sufficient for FHE workloads. With the architectures scalibility, it can achieve $2.4$TB/s with 192 channels—on par with the NVIDIA A100~\cite{choquette2021nvidia}—and up to $12.8$TB/s with 1024 channels. In contrast, electrical interconnects deliver only $5.26$GB/s at a latency of $3.04~\text{ns}$ with a 128-channel configuration. Even with 1024-bit data sequences, their bandwidth is limited to $42.1$GB/s.  Achieving \emph{OptoLink}'s $1.6$TB/s bandwidth electrically would require an unfeasible $4864$-bit data width.

The ultra-fast data transfer of \emph{OptoLink} minimizes latency between memory and computational units, alleviating bottlenecks in FHE accelerators. By enabling efficient data exchanges like coefficients and twiddle factors, it accelerates operations, supports large datasets, and enhances performance for privacy-preserving applications. Its scalability ensures suitability for evolving computational demands.

\begin{table}[htbp]
\caption{Bitrate comparison of Electrical Network and \emph{OptoLink}}
\begin{center}
\begin{tabular}{|c|c|c|c|c|}
\hline
 & \multicolumn{2}{|c|}{\textbf{Electrical Network}} & \multicolumn{2}{|c|}{\textbf{OptoLink}} \\ 
\cline{2-5} 
\textbf{Bitwidth} & \textbf{\textit{Latency}} & \textbf{\textit{Bitrate}} & \textbf{\textit{Latency}} & \textbf{\textit{Bitrate}} \\
\hline
32 & 3.04ns & 1.32GB/s & 10ps & 0.4TB/s \\
\hline
64 & 3.04ns & 2.63GB/s & 10ps & 0.8TB/s \\
\hline
128 & 3.04ns & 5.26GB/s & 10ps & 1.6TB/s \\
\hline
\end{tabular}
\label{tab:timing_delay}
\end{center}
\vspace{-0.25in}
\end{table}

\subsection{Power Analysis}\label{subsec:power}

The \emph{OptoLink} system's power consumption is largely dictated by its laser source, MRRs, and PDs. The total power consumption can be expressed as,
\begin{equation}
P_{\text{total}} = P_{\text{laser}} + P_{\text{TX}} + P_{\text{RX}},
\label{e4}
\end{equation}
where \( P_{\text{laser}} \) accounts for the laser source's power usage, while \( P_{\text{TX}} \) and \( P_{\text{RX}}\) represent the power consumed by the transmitter and receiver, respectively. Each transmitter includes MRR thermal heating, which consumes approximately $0.32~\text{mW}$ per resonator~\cite{joshi2009silicon}, resulting in \(P_{\text{TX}} \) $=1.22 mW$ and \(P_{\text{TX}} \) $=0.92 mW$ per optical channel. For a 128-channel \emph{OptoLink} system supporting 4 NTT cores, the estimated power consumption is $6.59~\text{W}$, scaling to $13.16~\text{W}$ for 8 cores and $26.31~\text{W}$ for 16 cores due to the additional transmitters and receivers required for increased core counts.

In contrast, electrical interconnects consume significantly less power. Under a 128-bit configuration, power consumption is $336.99~\mu\text{W}$ for 4 cores, $661.74~\mu\text{W}$ for 8 cores, and $1332.31~\mu\text{W}$ for 16 cores. For narrower 32-bit configurations, electrical networks consume between $283.89~\mu\text{W}$ and $1121.9~\mu\text{W}$, while \emph{OptoLink} requires $1.65~\text{W}$ to $6.58~\text{W}$. Similarly, under 64-bit configurations, electrical network power ranges from $308.18~\mu\text{W}$ to $1232.19~\mu\text{W}$, whereas \emph{OptoLink} consumes $3.29~\text{W}$ to $13.16~\text{W}$.


The increased power demand of \emph{OptoLink} comes from optical components, with lasers consuming significant energy for stable light, and transmitters/receivers adding overhead due to photodetectors and signal processing. MRRs also require thermal control, further raising power usage. In contrast, electrical interconnects are more power-efficient with simpler designs. Despite the higher power consumption, \emph{OptoLink} offers superior scalability and high data throughput, making it ideal for applications prioritizing performance and bandwidth over energy efficiency.

\begin{figure}[!t]
\centering
\includegraphics[width=0.48\textwidth]{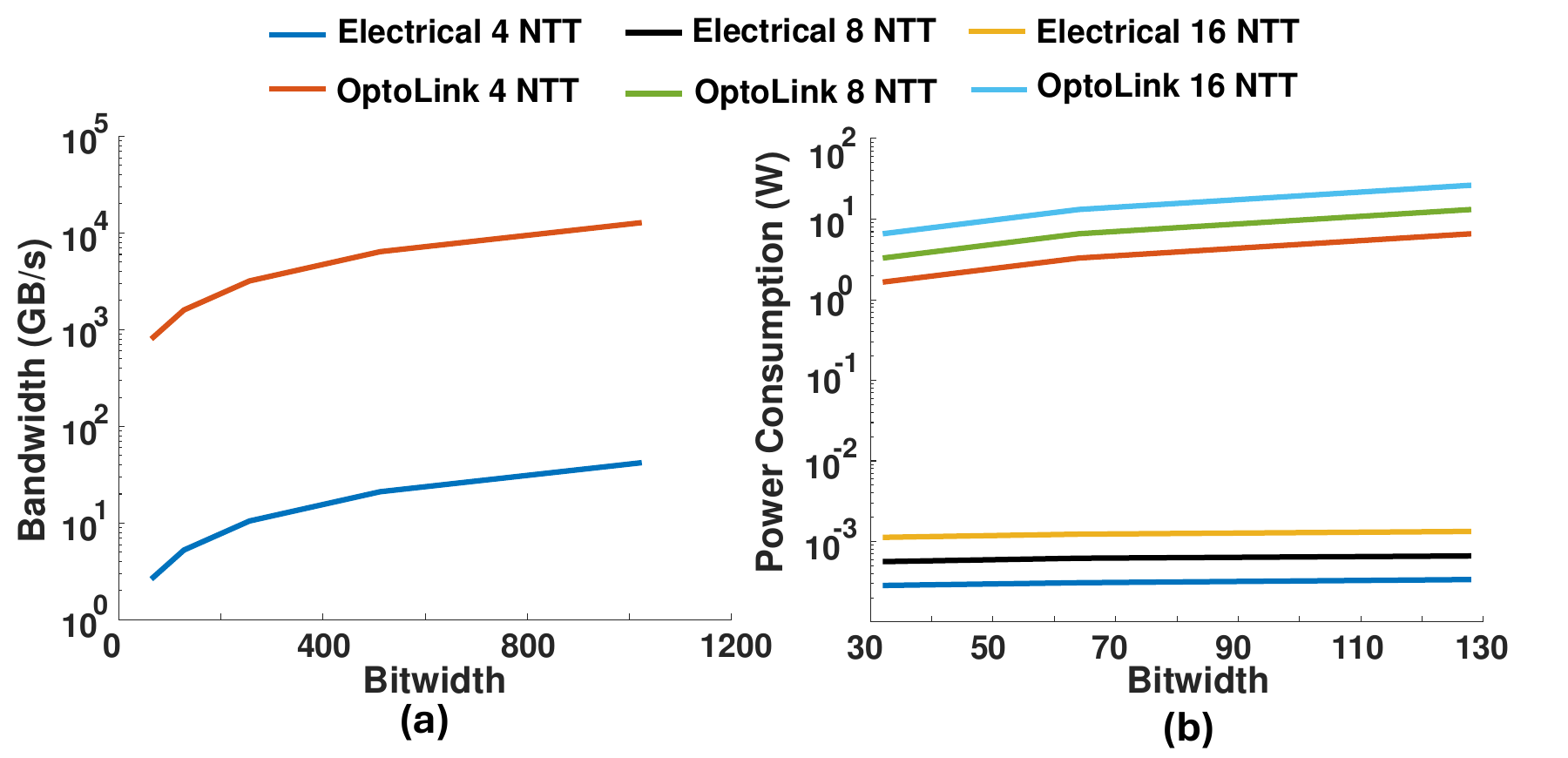}

\caption{{Comparison of electrical network and $OptoLink$ performance across varying numbers of NTT cores. (a) Relationship between bitwidth and bandwidth, and (b) Relationship between bitwidth and power consumption.}}
\label{fig:bits_vs_power}
\vspace{-0.2in}
\end{figure}

\begin{table}[htbp]
\caption{Power Consumption of Electrical Network and \emph{OptoLink}}
\begin{center}
\begin{tabular}{|c|c|c|c|}
\hline
\textbf{Bitwidth} & {\textbf{NTT Cores}} & \multicolumn{2}{|c|}{\textbf{Power Consumption}} \\ 
\cline{3-4} 
 & & \textbf{\textit{Electrical Network}} & \textbf{\textit{OptoLink}} \\
\hline
32 & 4 & 283.89 $\mu W$ & 1.65 $W$ \\
\cline{2-4} 
 & 8 & 562.44 $\mu W$ & 3.29 $W$ \\
\cline{2-4} 
 & 16 & 1121.9 $\mu W$ & 6.58 $W$ \\
\hline
64 & 4 & 308.18 $\mu W$ & 3.29 $W$ \\
\cline{2-4} 
 & 8 & 619.29 $\mu W$ & 6.58 $W$ \\
\cline{2-4} 
 & 16 & 1232.19 $\mu W$ & 13.16 $W$ \\
\hline
128 & 4 & 336.99 $\mu W$ & 6.59 $W$ \\
\cline{2-4} 
 & 8 & 661.74 $\mu W$ & 13.16 $W$ \\
\cline{2-4} 
 & 16 & 1332.31 $\mu W$ & 26.31 $W$ \\
\hline
\end{tabular}
\label{tab:power_consumption}
\end{center}
\vspace{-0.25in}
\end{table}

\subsection{Area Analysis}\label{subsec:area}

Comparing the space usage of traditional electronic networks with that of the photonic components essential to \emph{OptoLink} allowed for an analysis of the area needs of the suggested \emph{OptoLink} architecture. The electronic network areas were approximated using a $32~\text{nm}$ technology library and realistic process design parameters.  For a 128-bit NTT configuration, the area requirements of the electrical network scaled nearly linearly with the number of NTT units, with 4, 8, and 16 NTT units occupying $3097.3~\mu\text{m}^2$, $5741.2~\mu\text{m}^2$, and $11861.9~\mu\text{m}^2$, respectively. \emph{OptoLink}, on the other hand, needs more space because of its photonic elements. According to ~\cite{thonnart201810gb}, each photonic transmitter or receiver takes up around $0.0096~\text{mm}^2$ per wavelength, and the wavelength-selective MRRs add an extra $0.01~\text{mm}^2$ based on an MRR radius of $5~\mu\text{m}$~\cite{li201240}. Additionally, MRRs require electrical connections, which increase the overall footprint. These connections include four wires for data transfer and temperature tuning.

\section{Conclusion}\label{sec:conclusion}

The \emph{OptoLink} architecture addresses key challenges in FHE accelerators by using photonic interconnects for ultra-low latency and high bandwidth. With picosecond-scale latencies and $1.6~\text{TB/s}$ throughput in a 128-channel configuration, it handles large ciphertexts and complex memory access patterns, reducing bottlenecks in tasks like key-switching and NTTs. While photonic components increase power and area demands, these are outweighed by performance gains. Future integration of broadcast-enabled photonic devices and an optomized dataflow will further optimize power and area efficiency by reducing the number of wavelengths and waveguides needed, enhancing scalability and energy efficiency. In summary, \emph{OptoLink} is a high-performance, scalable interconnect solution tailored to the demands of FHE systems, enabling faster and more efficient privacy-preserving computations. Ongoing advancements will refine its efficiency, ensuring its broader applicability in data-intensive applications.




%


\bibliographystyle{IEEEtran}
\bibliography{IEEEabrv,GOMACTech_LaTeX}

\end{document}